\documentstyle[prl,aps,preprint]{revtex}
\begin{document}
\draft

\title{Coherent control of an atomic collision in a cavity }
\author{S. Osnaghi$^1$, P. Bertet$^1$, A. Auffeves$^1$, P. Maioli$^1$, M. Brune$^1$, 
J.M. Raimond$^1$ and S. Haroche$^{1,2}$}

\address{$^1$Laboratoire Kastler Brossel\\ D\'epartement de Physique de
l'Ecole Normale Sup\'erieure,\\ 24 rue Lhomond, F-75231 Paris
Cedex 05 France\\
$^2$ Coll\`ege de France, 11 place Marcelin-Berthelot, F-75005, Paris France}

\date{\today}
\maketitle

\begin{abstract}

Following a recent proposal by S. B. Zheng and G. C. Guo (Phys. Rev. Lett. {\bf 85}, 
2392 (2000)), we report an experiment in which two Rydberg atoms crossing 
a non-resonant cavity are entangled by coherent energy exchange. The process,  
mediated by the virtual emission and absorption of a microwave photon, is characterized 
by a collision mixing angle four orders of magnitude larger than for atoms 
colliding in free space with the same impact parameter. The final entangled state 
is controlled by adjusting the atom-cavity detuning. This procedure, 
essentially insensitive to thermal fields and to photon decay, opens promising 
perspectives for complex entanglement manipulations. 

\end{abstract}

\pacs{03.67-a, 42.50-p, 34.60+z}

Conditional dynamics and quantum gates \cite{GATES} involving individually addressable 
particles have been demonstrated in various quantum optics experiments. Beyond their 
fundamental interest to test basic aspects of quantum theory, these studies open new 
perspectives in quantum information processing \cite{QUANTINFO}. In most schemes, 
information is carried by internal degrees of freedom of atomic particles, called 
``qubits". Logic gates are realized via coherent ``collisions" between them. The qubits 
are put in contact on demand, coupled for a given time, then separated while the 
interaction with the environment, causing decoherence \cite{DECOHERENCE}, is kept to a minimum. 
Such processes are quite different from usual collisions whose output is determined 
only statistically. 

In ion trap experiments, the collision is achieved by establishing the contact between the qubits 
with lasers, through Raman processes involving the excitation of vibrational modes of motion of 
the ions \cite{IONGATE}. In cavity quantum electrodynamics (CQED), a primary collision involves 
an atom and a cavity mode in exact resonance. The two systems are entangled as the atom exits 
the cavity \cite{EPRPAIR,GATE}. Atom-atom entanglement is obtained by combining such atom-field 
collisions \cite{OURGHZ}. Other proposals involve collisions between cold atoms trapped by light, 
which are put in contact, then separated, by adiabatically changing the laser beam parameters 
\cite{ATOMS}. Several schemes involve the coupling between atoms momentarily excited into Rydberg 
states \cite{ZOLLERRYD}. Owing to their large electric dipoles, these states are ideal to achieve 
strong qubit interactions. Although it has not yet been used to build quantum gates, the van der 
Waals interaction between excited atoms has been investigated in the early days of Rydberg atom 
physics \cite{VITRANT} and recently revisited in the context of cold atom studies \cite{PILLET}. 

Following a recent proposal \cite{ZHENG}, we describe in this Letter an experiment in which 
we control the collision of two Rydberg atoms in a process assisted by a non-resonant cavity. 
The atoms exchange their energy and get entangled while they cross together the cavity. This 
process bears similarities with light-induced atomic collisions \cite{LIGHT}, with the 
difference that, in the present case, the field modes enhancing the collision rate are essentially 
empty (vacuum field effect). The cavity makes the entanglement process about $10^4$ times more 
efficient than a free space collision with the same impact parameter. The final atomic entangled 
state is tailored by adjusting the atom-cavity detuning. Contrary to previous CQED experiments 
\cite{EPRPAIR,GATE,OURGHZ}, this cavity-assisted entanglement process 
leaves the field unexcited and is essentially insensitive to thermal cavity excitations and to 
cavity losses. These features make this method very promising for quantum information processing 
\cite{QUANTINFO,ZHENG}. This cavity-assisted collison process can also be related to recently 
proposed \cite{MOLMER,ZAGURY} and implemented \cite{IONFOUR} schemes  in ion trap physics, in 
which atomic entanglement is realized via virtual vibrational excitations of the ions. 

Let us first recall some orders of magnitude relevant to van der Waals collisions between Rydberg 
atoms in free space. We consider the resonant energy exchange $|e_1,g_2\rangle\rightarrow|g_1,e_2\rangle$
between two atoms $A_1$ and $A_2$ initially in states $e$ and $g$ respectively ($e$ and $g$ correspond 
to large principal quantum numbers $n$ and $n-1$). The atoms interact via the dipole-dipole coupling \linebreak
$W_{vdW}(R,{\mathbf{u}})=[{\mathbf{r}}_1\cdot{\mathbf{r}}_2-3({\mathbf{r}}_1\cdot{\mathbf{u}})
({\mathbf{r}}_2\cdot{\mathbf{u}})]q^2/4\pi\varepsilon_0R^3$, where ${\mathbf{r}}_i\ (i = 1,2)$ 
are the valence electron (charge $q$) coordinates 
in each atom, $R$ and $\mathbf u$ the distance between the atoms and the unit vector along the interatomic 
direction respectively. The expression of $W_{vdW}$, simply derived from electrostatic laws, can also be 
viewed as resulting from virtual photon exchange between the atoms, summed over all possible field modes. 
A collision is characterized by the functions of time $R(t)$ and ${\mathbf{u}}(t)$ which determine the 
``collision mixing angle" $\theta=(1/\hbar)\int\,dt\,|\langle e_1,g_2|W_{vdW}(R(t),{\mathbf{u}}(t))|g_1,e_2\rangle|$. 
The atoms emerge from the collision in the generally entangled state~: 
\begin{equation}
|\Psi\rangle=\cos\theta|e_1,g_2\rangle+\exp(i\Phi)\sin\theta|g_1,e_2\rangle
\end{equation}
where $\Phi$ is a phase depending on $R(t)$ and ${\mathbf{u}}(t)$ which we need not specify.

Even in free space, the large-sized Rydberg atoms are very sensitive to the van der Waals 
interaction. For a collision with an impact parameter $b_0$, involving atoms with relative 
velocity $v$, an order of magnitude estimate yields~: 
\begin{equation}
\theta_0=n^4\,\frac{q^2}{4\pi\varepsilon_0\hbar v}\,\frac{a_0^2}{b_0^2}=\alpha\,\frac{c}{v}\,\left(\frac{a_0n^2}{b_0}\right)^2
\end{equation}
where $a_0 = 0.53\ 10^{-10}$ m is the Bohr radius and $\alpha=q^2/4\pi\varepsilon_0\hbar c=1/137$ 
the fine structure constant. For $n = 51$ and $v/c=10^{-6}$  (typical atomic beam velocity), the 
condition $\theta_0=\pi/4$  of maximum entanglement is achieved for $b_0=13\ \mu$m,  a huge distance 
at atomic scale. We show in this Letter that, by having the atoms interact not in free space, but 
in a cavity, the Rydberg-Rydberg collision angle is enhanced by a huge factor, making it possible 
for the atoms to get entangled while they collide at millimetric distances. 

Our set-up is shown in Fig. 1(a). The Rb atoms, effusing from an oven $O$, propagate along an 
horizontal beam crossing the cavity $C$ made of two superconducting niobium spherical mirrors placed 
at $L= 2.75$~cm from each other \cite{EPRPAIR,GATE,OURGHZ,QND}. The set-up is cooled to 1.3 K to minimize 
thermal radiation. The atoms are velocity selected by laser optical pumping, according to a procedure 
described elsewhere \cite{EPRPAIR}. They are then prepared in box $B$ by a combination of laser and 
radiofrequency excitation in the circular Rydberg states with principal quantum numbers $n = 51$ ($e$) 
or 50 ($g$). The atoms, prepared with different velocities, collide inside $C$ which they cross at 
mid-distance between the mirrors. After exiting the cavity,  they are detected by a state selective 
field ionization detector $D$ (efficiency 40\%), discriminating with less than $5$\% error rate $e$ 
and $g$. An optional classical microwave pulse $R$ coherently mixing $e$ and $g$ ($\pi/2$ pulse) can 
be applied to the atoms after $C$, for analyzing the final state of the collision process. The sequence 
of events is schematized in Fig. 1(b) which shows a space-time diagram depicting the evolution of the 
two atoms crossing the apparatus. 

The Rydberg excitation is pulsed within a time of 2 $\mu$s. In each atomic pulse, we prepare on the 
average 0.25 atom, with Poisson statistics. The probabilities for exciting 0, 1 and 2 atoms per pulse 
are respectively 0.78, 0.19 and 0.025. Events in which only one atom is detected in the two pulses 
(0.6\% of the experimental sequences) are recorded. In approximately 25\% of these events, there are 
in fact two atoms in one of the pulses, one of them escaping detection. These ``three atom collision" 
events are a source of errors. We focus here on the case where a simple $A_1-A_2$ pair has been prepared. 
We choose the delay $T$ between the preparation pulses and the two atomic velocities $v_1$ , $v_2$ such 
that $A_1$ overcomes $A_2$ at cavity center. This event defines the time origin $t=0$. The $A_1-A_2$ 
separation at $t=0$ is of the order of the atomic beam diameter, about 0.5 mm. This would be the impact 
parameter for the same collision process in the absence of the cavity, corresponding to a negligible 
entanglement ($\theta_0\simeq5.10^{-4}$)    

The cavity sustains two $TEM_{900}$ modes, $M_a$ and $M_b$, with linear orthogonal polarizations and 
transverse gaussian profiles (common waist at center $w = 0.6$ cm). Because of a small mirror anisotropy, 
the mode degeneracy is lifted (frequency difference $\Delta/2\pi = 128$ kHz). The two frequencies $\omega_a$ 
and $\omega_b=\omega_a+\Delta$  can be tuned together by translating the mirrors with a piezostack. The 
modes are frequency shifted from the atomic $e\rightarrow g$ transition frequency $\omega/2\pi = 51.1$ GHz 
by variable detunings $\delta_a$ and $\delta_b=\delta_a+\Delta$.  The field mode damping times are $T_{c,a}=10^{-3}$ 
s and $T_{c,b}=0.9\ 10^{-3}$ s. The maximum vacuum field r.m.s. amplitude in each mode is 
$E_0=(2\hbar\omega/\pi\varepsilon_0Lw^2)^{1/2}=1.57\,10^{-3}$~V/m.  At equilibrium, there is an average of about 
one thermal photon per mode, due to microwave leaks in $C$. These photons are erased at the beginning of each 
experimental sequence by sending a train of absorbing atoms across $C$ \cite{QND}. During the 180 $\mu$s delay 
between the end of the erasing sequence and the $A_1-A_2$ collision, a field of about 0.25 photon builds up in 
each mode. 

In a first experiment, the $R$-pulse is not used. The delay between the atomic preparations is $T = 78\ \mu$s, 
with $v_1= 300$ m/s , $v_2 =243$ m/s. We sweep $\delta_a$ (and $\delta_b=\delta_a+\Delta$). For each detuning 
value, we detect 1000 atomic pairs and we reconstruct the four detection probabilities  
$P(e_1,g_2)$, $P(g_1,e_2)$, $P(e_1,e_2)$ and $P(g_1,g_2)$. Fig. 2 shows the variations of these 
probabilities as a function of the dimensionless 
detuning parameter  $\eta=(\omega/\delta_a+\omega/\delta_b)$. We see that $P(e_1,g_2)$ and $P(g_1,e_2)$ (solid and 
open circles respectively) oscillate in a symmetrical way as a function of $\eta$. These variations reflect the 
pattern described by Eq.~(1), the detuning parameter $\eta$ being - as we show below - directly related to the 
cavity-assisted collision angle. The other probabilities $P(e_1,e_2)$ and $P(g_1,g_2)$ are due to erroneous 
detection counts and to three-atom collisions. They remain on the average at a low background level (about $10$\%). 
To get a ``zero-collision angle" reference, we have changed $T$ to 108~$\mu$s, so that the atoms now crossed 37 mm 
downstream the cavity axis and set $\delta/2\pi=470$ kHz. We measured then $P(g_1,e_2)= 0,01\ (\pm 0.01)$ and 
$P(e_1,g_2) = 0.89\ (\pm 0.01)$ instead of the ideal 1 value, due to detection errors. This demonstrates that 
the collision effect observed here is fully cavity-assisted as discussed above. The corresponding experimental 
points have been put in Fig. 2 at $\eta=0$ (equivalent to ``infinite" cavity detuning).  

The coupling of $A_1$ and $A_2$ to each mode depends upon atomic positions. At cavity center ($t =0$), this 
coupling is characterized by the Rabi frequency $\Omega=2D_{eg}\cdot E_0/\hbar$  where $D_{eg}=qa_0n^2/2$ 
is the dipole matrix element between $e$ and $g$  ($\Omega/2\pi=50$ kHz deduced from atomic and cavity 
parameters, to be compared with the experimental value $\Omega/2\pi=49\pm 1$~kHz \cite{QRABI}). At time $t$, 
$A_1$ and $A_2$ have moved away from cavity center and their coupling to each cavity mode is 
$\Omega\exp(-v_i^2t^2/w^2)$, $(i=1,2)$. The energy exchange process involves the virtual emission by the 
first atom of one photon in one mode ($|e_1,g_2;0_\mu\rangle\rightarrow|g_1,g_2;1_\mu\rangle$; $\mu=(a,b)$, 
combined to the photon absorption by the second atom 
($|g_1,g_2;1_\mu\rangle\rightarrow|g_1,e_2;0_\mu\rangle$). It is easy to compute the effect of such virtual 
transitions in the limit $\delta_a\gg\Omega$  (non-resonant cavity QED regime).  The contribution of the 
intermediate state $|g_1,g_2;1_\mu\rangle$ to the mixing angle involves the product of the atom's couplings 
divided by the frequency mismatch $\delta_a$. Summing over the modes and averaging the variations of 
the atom-cavity coupling, we find the cavity-assisted collision mixing angle~:
\begin{equation}
\theta_c=\Omega^2\,\left(\frac{1}{\delta_a}+\frac{1}{\delta_b}\right)\,\frac{\sqrt\pi}{4\sqrt 2}\,\frac{w}{v_0}\ ,
\end{equation}        
with $v_0=[(v_1^2+v_2^2)/2]^{1/2}$.                    

Finally, replacing in Eq. (3) $\Omega$ by its expression in terms of $D_{eg}$ and $E_0$ and 
these quantities by their expressions in terms of cavity and atom parameters, we find:
\begin{equation}
\theta_c=\alpha\,\left(\frac{\omega}{\delta_a}+\frac{\omega}{\delta_b} \right)\,\frac{c}{v_0}\left( \frac{a_0n^2}{b_c}\right)^2\ ,
\end{equation} 
with $b_c=(Lw/\sqrt{2\pi})^{1/2}=0.81$ cm.

While Eq. (2) is qualitative, Eq. (4) is exact for $\delta_a\gg\Omega$. Comparing Eq. (4) and (2), 
we find that $\theta_c$ is obtained by multiplying by $\eta$ the free space collision angle 
corresponding to an impact parameter $b_0=b_c$. Since $b_c$ is about three orders of magnitude 
larger than the $b_0$ value corresponding to $\theta_0=\pi/4$  in free space, maximum  entanglement 
in the cavity is achieved with $\eta$ of the order of $10^6$. The solid lines in Fig. (2) are 
obtained by replacing in Eq.~(1) $\theta_0$ by $\theta_c$ and by multiplying $P(e_1,g_2)$ and 
$P(g_1,e_2)$ by 0.89, in order to fit the data at $\eta=0$ and thus account for detection errors. 
We note a good agreement between the experiment and this simple model for $\eta<5.\,10^5$ , i.e. $\delta_a>3\Omega$. 

In this perturbative regime, the collision is to first order insensitive to thermal photons. If 
there are $N_\mu$ photons in one mode, the virtual process in which an additional photon is 
emitted  ($|e_1,g_2;N_\mu\rangle\rightarrow|g_1,g_2; N_\mu+1\rangle\rightarrow|g_1,e_2;N_\mu\rangle$) 
interferes destructively with the one in which a photon is absorbed 
($|e_1,g_2;N_\mu\rangle\rightarrow|e_1,e_2;N_\mu-1\rangle\rightarrow|g_1,e_2; N_\mu\rangle$), 
since the corresponding amplitudes have opposite signs. The net result is $N_\mu$-independent 
and identical to the one obtained for a cavity in its vacuum state. This is verified by 
observing that the solid line theoretical curve, computed by assuming  $N_\mu=0$, fits 
well the results of the experiment, in which the probability to have $N_\mu=1$ is 0.25. 
A similar insensitivity to thermal excitations of atom-atom interaction mediated by virtual 
coupling to a vibration mode occurs in ion traps \cite{MOLMER,ZAGURY}.  

For larger $\eta$ values the condition $\delta_a/\Omega\gg 1$ is no longer valid and the 
collision angle departs from the perturbative expression (Eq. (4)), even for cavity modes 
at zero temperature. In addition, the effect of thermal field excitations cannot then be 
neglected. Thus, we numerically solve the equations of motion of the 
two atoms in the cavity, taking into account exact atom-cavity coupling as well as the 
0.25 thermal photons per mode (but neglecting cavity relaxation during interaction time). 
The theory is normalized to fit the data at $\eta=0$. The results are given by the dotted 
curves in Fig. (2), which reproduce qualitatively well the variations of $P(e_1,g_2)$ and 
$P(g_1,e_2)$ in the whole range of $\eta$ values up to a $2\pi$ collision angle. Note however that, 
for large $\eta$ values, the contrast of the experimental oscillations is smaller than the 
theoretical one. Part of this contrast reduction originates in three-atom collision processes. 
For small $\eta$'s, up to the point $\theta_c=\pi/4$  of maximum entanglement, the two-atom 
collision model is quite satisfactory.

The coherence of the cavity-assisted collision is checked in a second experiment. We choose a 
different set of parameters~: $T=115\ \mu$s, $v_1=500$ m/s, $v_2=319 $ m/s. Fixing $\eta$ to 
realize $\theta_c=\pi/4$, we apply independently to $A_1$ and $A_2$ a $\pi/2$ pulse $R$ 
(with a frequency $\omega_r$ close to $\omega$), realizing a basis change in the $e-g$ subspace. 
The delay between the pulses is $\tau=22\ \mu$s. In the Bloch vector representation, the 
$|e\rangle$ and $|g\rangle$ states correspond to a ``pseudo-spin" along the ``$Oz$ axis". 
Detecting the energy of $A_1$ after the $R$ pulse amounts to a ``transverse" detection of the 
corresponding pseudo-spin. Finding $A_1$ in $e$ (resp. $g$) is then equivalent to measuring 
it along the ``$Ox$ axis" (eigenstate $|+_x\rangle$ of the Pauli matrix $\sigma_x$ (resp. 
$|-_x\rangle$)). For $A_2$, we detect in the same way the states $|\pm_\phi\rangle$ and,
 eigenstates of $\sigma_\phi=\cos\phi\,\sigma_x+\sin\phi\,\sigma_y$, with 
$\phi=(\omega-\omega_r)\tau$. 

By repeating the experiment while sweeping $\omega_r$ (thus $\phi$), we reconstruct the 
combination of joint probabilities  
$P(+_{1,x};+_{2,\phi})+P(-_{1,x};-_{2,\phi})-P(+_{1,x};-_{2,\phi})-
P(-_{1,x};+_{2,\phi})=\langle\sigma_{1,x}\,\sigma_{2,\phi}\rangle$. This ``Bell signal" , 
shown in Fig. 3 versus $\phi$, measures the angular correlations between the transverse 
spin components associated to the two atoms.  Ideally, the process should prepare a pair 
of maximally entangled EPR particles, with a signal oscillating between $+ 1$ and $-1$. 
The reduced contrast of the observed modulation, about 50\%, is due to the already 
mentioned defects in the entangled state preparation, as well as imperfections in the $R$ pulses.   

After we have improved our set-up (notably by preparing the atoms via a deterministic and 
not a Poissonian process), many promising experiments generalizing the present study will 
become possible. By combining a two-atom cavity assisted collision with single atom 
unitary operations, robust quantum gates directly coupling atomic qubits could be realized \cite{ZHENG}
and new tests of Bell's inequalities with atoms performed. Situations where three atoms at
 a time cross the cavity and interact with its field via real or virtual photon processes 
could lead to the realization of useful three-bit logical gates.

{\sl Acknowledgements}~: Laboratoire Kastler Brossel, Universit\'e Pierre
et Marie Curie et ENS, associ\'e au CNRS (UMR 8552). We acknowledge
support of the European Community and of the Japan Science and Technology
corporation (International Cooperative Research Project~: Quantum Entanglement).

\begin{figure}
\caption{ (a) Scheme of the experimental apparatus. (b) Space-time diagram depicting 
the sequence of events. Atoms $A_1$ and $A_2$, sent at different times with velocities 
$v_1$ and $v_2$, simultaneously cross the cavity axis at time $t=0$. They undergo an 
optional microwave pulse $R$ before being detected by field ionization in $D$.}
\end{figure}

\begin{figure}
\caption{Joint detection probabilities versus the detuning parameter $\eta$. $P(e_1,g_2)$ 
and $P(g_1,e_2)$ (solid and open circles) oscillate in a symmetrical way, reflecting the 
atom-atom energy exchange enhanced by the cavity. The solid line represents the 
predictions of Eq. (4), in the $\eta<5\,10^5$ range where it applies. The dashed lines 
present the result of a numerical integration of the system's evolution. The spurious 
channels probabilities, $P(e_1,e_2)$ and $P(g_1,g_2)$ (open squares and diamonds respectively) stay below the 10\% level.}
\end{figure}

\begin{figure}
\caption{``Transverse" correlations~: Bell signal $\langle\sigma_{1,x}\,\sigma_{2,\phi}\rangle$ versus relative phase $\phi$.
The modulation reveals the coherence of the cavity-assisted collision process}
\end{figure}

\end{document}